\documentclass{elsart}
\usepackage{amssymb}
\usepackage{graphicx}
\usepackage{epsfig}

\def\beq{\begin{equation}}
\def\eeq{\end{equation}}
\def\bqa{\begin{eqnarray}}
\def\eqa{\end{eqnarray}}
\def\Gwp{\Gamma_{\mu}^{(D)}(p, p^{\prime}, Q)}
\def\pp{p^{\prime}}
\def\Gwx{\Gamma_{\mu}^{(D)}(x, y)}
\def\l0{\langle 0 |}
\def\r0{| 0 \rangle}
\def\qq{\langle\bar{q}q\rangle}
\def\Su{S_{u,d}}
\def\Sc{S_c}
\def\D{\mathcal{D}}
\def\B{\mathcal{B}}
\def\rw{\rightarrow}
\def\Bw#1{B_{\mu}^{(#1)}}

\begin{document}

\begin{frontmatter}

\title{The $\omega DD$ vertex in a Sum Rule approach}

\author[a1]{L.\ B.\ Holanda} 
\author[a2]{R.\ S.\ Marques de Carvalho}
\author[a3]{A.\ Mihara}
\ead{mihara@if.sc.usp.br}

\address[a1]{Instituto de F\'\i sica Te\'orica, UNESP     \\ 
R.\ Pamplona, 145, 01405-900 S\~ao Paulo, SP - Brazil} 
\address[a2]{Campus Experimental de Itapeva, UNESP      \\ 
R.\ Geraldo Alckmin, 519, 18409-010, Itapeva, SP - Brazil}
\address[a3]{Instituto de F\'\i sica de S.\ Carlos, USP   \\ 
C.\ P.\ 369, 13560-970 S\~ao Carlos, SP - Brazil} 

\begin{abstract}

The study of charmonium dissociation in heavy ion collisions
is generally performed in the framework of effective Lagrangians
with meson exchange. Some studies are also developed with the 
intention of calculate form factors and coupling constants related 
with charmed and light mesons. These quantities are important in 
the evaluation of charmonium cross sections.
In this paper we present a calculation of the $\omega DD$ vertex
that is a possible interaction vertex in some meson-exchange models
spread in the literature.  
We used the standard method of QCD Sum Rules in order to obtain the
vertex form factor as a function of the transferred momentum.      
Our results are compatible with the value of this vertex form factor
(at zero momentum transfer) obtained in the vector-meson dominance
model.                     
   
\end{abstract}

\end{frontmatter}


\section{Introduction}
\noindent
 
One of the most relevant topic 
in the physics of relativistic heavy
ion collisions is the interaction of charmonium with nuclear matter,
since the charmonium suppression is one of the most evident signals
of formation of the quark-gluon plasma (QGP) \cite{matsui}. 
The $J/\psi$ production and absorption in hadronic matter is still an open 
theoretical discussion \cite{muller}. It is important to understand the 
two different ways of charmonium absorption: by the nucleons and by the 
co-mover light mesons ($\pi, K, \rho, \omega$, etc...). Calculations 
of form factors and coupling constants related with charmed and light 
mesons are of great importance in evaluating charmonium cross sections
\cite{finpe}. Recent results of BABAR, CLEO, BELLE and SELEX 
on $D$ mesons spectroscopy
stimulate calculations of physical quantities like vertices involving $D$ 
mesons \cite{bianco}.

In the low energy regime ($\sqrt{s}\lesssim$ 10GeV), theoretical studies 
are performed mainly through the use of effective Lagrangians. In 
this context, one has a good control of the relevant symmetries that 
underly the dynamics of the process. On the other hand, it is necessary 
to know the values of form factors -- associated with the vertices -- 
with some precision. The choice of a lower or higher value of the form 
factor may change the final cross section in some orders of magnitude.

In refs.~\cite{muller1,haglin,lin-ko} effective models for $J/\psi$ 
absorption in hadronic matter are proposed. 
The Lagrangians involve charmed and light 
pseudoscalar mesons (P) and also vector mesons (V). The interaction among 
these particles occurs through three--point (PPV and VVV) and
four--point (PPVV and VVVV) vertices. 
One possible vertex in some of these models involves one $\omega$ and two 
$D$ mesons. On the other hand,
in the framework of vector meson dominance (VMD) model, one can obtain the value 
of the coupling constant (at zero momentum transfer)
for such a vertex as $g_{\omega DD}\approx -2.84$ 
(Appendix A of \cite{lin-ko} and also \cite{klingl}).

The QCD sum rules (QCDSR) method \cite{svz} have been used in many works 
to calculate cross-sections, form factors and coupling constants 
(see for instance \cite{yang,reinders,ioffe1,ioffe2,revqcdsr}). 
More recently, other works have used this method to obtain quantities 
related to charmonium suppression \cite{jpsi,rho}. 
These works stimulate the use of QCDSR in these kind of problems.
Some works have used form factors calculated with QCDSR method
for charmed and light mesons \cite{finpe} to study the $J/\psi$ dissociation. 
These form factors depend on the transferred momentum and this dependence is taken 
in to accout in the calculation of cross sections.

Following these ideas, we present in this paper a study of $\omega DD$ 
vertex based on the QCDSR. This technique allows one to 
obtain hadronic quantities in terms of quark and gluon properties. 
Our results are compatible, within error estimates, with the value 
obtained through the VMD model. 

We follow the standard procedure of QCDSR. We calculate
the Operator Product Expansion (OPE) and the phenomenological 
contributions for the correlation function of $\omega DD$ vertex
and we equate both contributions following the principle of quark--hadron
duality. In order to suppress higher order contributions from
the OPE side as well as higher resonances (and continuum) from the 
phenomenological side, we use the Borel transform in both sides of the 
equation, obtaining the sum rule. 

We perform the numerical integration of the sum rule to
estimate the coupling constant. This coupling
constant is a function not only of the transferred momentum
$Q^2$ but also of the Borel masses. In general one 
considers the dependence of decay constants ($f_D$ and $f_{\omega}$) 
with Borel mass to improve the stability
of the coupling constant with respect to the variation of the Borel masses.        

\begin{figure}
{\centering 
\resizebox*{0.3\textheight}{!}
{\includegraphics{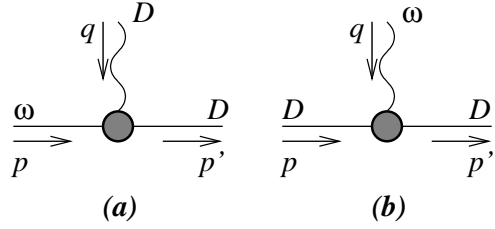}} 
\par}
\caption{The $\omega DD$ vertex, with mesons momenta, in two 
situations: $(a)$ $D$ off--shell and $(b)$ $\omega$ off--shell .}
\label{fig:vtx0}
\end{figure}

\section{Sum Rules for the vertex}

{\bf Vertex with $D$ off-shell.}
We start considering the $\omega DD$ vertex with one of the 
$D$ mesons off-shell (fig.~\ref{fig:vtx0}-$a$).
The three-point correlator is given by 
\begin{eqnarray}
\Gwp 
= \int d^4x \, d^4y\, e^{i \pp x} e^{-i q y}
\Gwx \,\, ,
\label{correl}
\end{eqnarray}
where
$\Gwx =
\l0 \mbox{T} \{ j_D(x)j^{\dagger}_D(y)j^{\dagger}_{\mu}(0) \} \r0 
$;
with the currents given by  $j_D = i \bar{u} \gamma_5 c$ for
$D$ and $j_{\mu}= \bar{u} \gamma_{\mu} u$ for $\omega$.

In the QCD side we perform the operator product expansion (OPE) \cite{muta} 
of the correlator $\Gwx$
\begin{eqnarray}
\Gwx = 
\bar{A}_{\mu} . 1 
+ 
\bar{B}_{\mu} . \qq + ...
\end{eqnarray}
where 1 is the identity operator, $\bar{A}_{\mu}(x,y)$ is the perturbative 
contribution, $\qq$ is the quark condensate and 
$\bar{B}_{\mu}(x,y)$ is the respective coefficient. Next, we present 
the calculations of $\bar{A}_{\mu}(x,y)$ and $\bar{B}_{\mu}(x,y)$ 
with some detail.

\begin{eqnarray}
A_{\mu}(x,y) = 
\mbox{Tr}\left[ \Su(-y)\gamma_5\Sc(y-x)
\gamma_5\Su(x-0)\gamma_{\mu} \right] ~,
\end{eqnarray}
where $\Su$ is the light quark $u$ (or $d$) propagator and $\Sc$
is the quark $c$ propagator. After Fourier transforming the propagators, 
defining the variables: $s=p^2$, $u=\pp\,^2$, $t=q^2$, using
Cutkosky rules and some extensive calculation 
we get the dispersion relation\footnote{Notation: 
$A_{\mu} = \bar{A}_{\mu} \cdot 1 $.}

\begin{eqnarray}
A_{\mu}(p, \pp, q)
= 
-\frac{1}{4\pi^2}\,\int_{s_{min}}^{\infty} ds \int_{u_{min}}^{\infty} du
\frac{\D_{\mu}^{(D)}}{(s-p^2)(u- p^{\prime 2})} ~,
\label{eq:Aw}
\end{eqnarray}
with the double discontinuity written as
\begin{equation}
\D_{\mu}^{(D)} = \rho\, p_{\mu} + \rho^{\prime}\, \pp_{\mu}~,
\end{equation}

and the spectral densities given by:
\begin{eqnarray}
\rho = 
-\frac{3}{2} \frac{(u-m_c^2)}{\sqrt{\lambda(s,t,u)}} -
\frac{6}{\pi}F 
\, , \,\,
\rho^{\prime} = 
\,\,\frac{3}{2} \frac{s}{\sqrt{\lambda(s,t,u)}} -
\frac{6}{\pi}G~,
\end{eqnarray}
with
\begin{eqnarray}
F &=&
\frac{\pi}{8}\frac{(s-t-u +2m_c^2)}{\sqrt{\lambda(s,t,u)}}\left[ 1 
- 
\frac{(s-t-u+2m_c^2)(s-t+u)}{\lambda(s,t,u)} \right], 
\nonumber\\
G&=&\frac{\pi}{4} \frac{s(s-t-u+2m_c^2)^2}{\lambda^{3/2}(s,t,u)}\,\, ,
\end{eqnarray}
where $\lambda$ is the kinematical function: 
$\lambda(x,y,z) =  (x-y-z)^2 - 4yz$

The integration limits of eq.~(\ref{eq:Aw}) ($s_{min}$ and $u_{min}$)
are determined from the conditions:
\begin{eqnarray}
s \geq 0 \, , \quad
u &\geq& t \nonumber\\
(s-t-u+2m_c^2)^2 &\leq& \lambda(s,t,u) \,\, .
\end{eqnarray}

We go to the Euclidean space with the transformations:
$p^2 \rw -P^2$, $p^{\prime 2} \rw -P^{\prime 2}$ and
$q^2 = t \rw -Q^2$. 
And after performing the double Borel transform in $P$ and 
$P^{\prime}$, we obtain:
\begin{eqnarray}
\B_{P^2}\left[\B_{P^{\prime 2}} [A_{\mu}] \right]
= 
-\frac{1}{4\pi^2}\,\int_{s_{min}}^{\infty} ds \int_{u_{min}}^{\infty} du \,
\D_{\mu}^{(D)}\, e^{-s/M^2}e^{-u/M^{\prime 2}}~.
\label{eq:pert1}
\end{eqnarray}

Now we consider the non-perturbative contributions. Here 
we shall concentrate only in the quark condensate contribution. 
The first contribution is \footnote{Notation: 
$B_{\mu} = \bar{B}_{\mu} \cdot \qq $.}:
\begin{eqnarray}
\Bw1 (p,\pp,q)= 
\int d^4xd^4ye^{i \pp x}e^{-i q y}
\Bw1(x,y)~,
\end{eqnarray}
where

\begin{eqnarray}
\Bw1(x,y)=\mbox{Tr}\left[ \left(-\frac{1}{12}\qq\right)~.
\gamma_5 
\Sc(y-x)\gamma_5\Su(x)\gamma_{\mu} \right]\,  ,\,\,   
\end{eqnarray}
After some trivial manipulation we get
 
\begin{equation}
\Bw1 (p,\pp,q) = \frac{m_c \, p_{\mu}}{p^2 (q^2 - m_c^2)}\,\qq~.
\end{equation}
It is not difficult to see that after going to the Euclidean 
space and taking the double Borel transform this contribution vanishes.

The second contribution is
\begin{eqnarray}
\Bw2(x,y) = 
\mbox{Tr}\left[\Su(-y)\gamma_5\Sc(y-x) ~,
\gamma_5\left(-\frac{1}{12}\qq\right)\gamma_{\mu}\right]  ~, 
\end{eqnarray}
that results in
\begin{equation}
\Bw2 (p,\pp,q) = \frac{- m_c \, p_{\mu}}{p^2 (p^{\prime 2} - m_c^2)}\,\qq~.
\end{equation}

Going to the Euclidean space and taking the double Borel transform we get
\begin{eqnarray}
\B_{P^2}\left[\B_{P^{\prime 2}} [\Bw2] \right]
= 
-m_c\qq 
e^{-m_c^2/M^{\prime 2}}p_{\mu} \, . 
\label{eq:npert1}
\end{eqnarray}
There is also a numerically negligible contribution from the charm 
condensate, which we will not take into account in this calculation.

After the double Borel transformation the correlator (\ref{correl}), in the
phenomenological side, can be written as  
\begin{eqnarray}
&&\B_{P^2}\left[\B_{P^{\prime 2}} [\Gwp|_{phen} ] \right]=
- \left( \frac{m_D^2 f_D}{m_c} \right)^2 \,m_{\omega}\,f_{\omega}
\,\frac{g_{\omega DD}^{(D)}(Q^2)}{m_D^2 + Q^2}\nonumber\\
&&\times e^{-m_{\omega}^2/M^2}e^{-m_D^2/M^{\prime 2}}
\left( -2\pp_{\mu}+
\frac{m_D^2 + m_{\omega}^2 + Q^2}{m_{\omega}^2}p_{\mu}\right)
+ \B_{P^2}\left[\B_{P^{\prime 2}} [\mbox{HR}] \right] ~,
\label{eq:pheno1}
\end{eqnarray}
where $g_{\omega DD}^{(D)}$ is the $\omega DD$ vertex, $m_{\mbox{M}}$ and 
$f_{\mbox{M}}$  are, respectively, the mass and the decay constant
of meson M ($\omega$ or $D$). HR represents the higher masses (continuum) 
resonances, whose contribution will be discussed below. 


One important thing to be considered is the model to be adopted 
for the spectral density  $\rho_{\mu}^{HR}(s,t,u)$ function (HR term in
the phenomenological side). We should remember that the interpolating 
fields couple not only with the fundamental state but also with all 
particles with the same quantum numbers. We should also consider 
the contribution of those states in the phenomenological side.
Admitting the quark--hadron duality, one assumes in general 
\begin{equation}
\rho_{\mu} ^{(HR)}(s,t,u) \approx  
\rho_{\mu} ^{(OPE)}(s,t,u) \, \theta(s - s_0 ) \, \theta(u - u_0 )
\quad ,
\end{equation}
where $s_0$ and $u_0$ are the continuum thresholds.

Proceeding this way, we have after the Borel 
transformation:
\begin{eqnarray}
\B_{P^2}\left[\B_{P^{\prime 2}} [HR] \right]
= 
-\frac{1}{4\pi^2}\,\int_{s_0}^{\infty} ds \int_{u_0}^{\infty} du \,
\rho_{\mu} ^{(OPE)}(s,t,u) \,e^{-s/M^2}\,e^{-u/M^{\prime 2}}~.
\label{eq:HR}
\end{eqnarray}

In the case under consideration, it is not difficult to see that
$\rho_{\mu} ^{(OPE)}(s,t,u) \rw \D_{\mu}^{(D)} (s,t,u)$.
Now we have all ingredients for computing of
$g_{\omega DD}^{(D)}(Q^2)$ \footnote{The superscript $(D)$ indicates
that $D$ is off-shell.} from the sum rule: 
(\ref{eq:pert1}) + (\ref{eq:npert1}) = (\ref{eq:pheno1}).

{\bf Vertex with $\omega$ off-shell.}
Now let us consider the case when the $\omega$ meson is
off-shell (fig.~\ref{fig:vtx0}-$b$). We follow a similar procedure 
of the previous section. The three-point correlator is written as
\beq
\Gamma_{\mu}^{(\omega)}(x, y) =
\l0 T{j_D(x)\, j^{\dagger}_{\mu}(y) \, j^{\dagger}_D(0)} \r0 \,\, .
\eeq 

For the perturbative contribution $A_{\mu}^{(\omega)}$, we obtained 
the double discontinuity $\D_{\mu}^{(\omega)}(s,u,t) = 
\rho(s,t,u).p_{\mu}  +  \rho^{\prime}(s,t,u)\pp_{\mu}$, with the
the spectral densities:
\bqa
\rho(s,t,u) &=& +3 \, \frac{(s-u-t)}{\lambda^{3/2}(s,u,t)}\,\, .
[m_c^2 (m_c^2 - s - u + t) + su] \\
\rho^{\prime}(s,t,u) &=& -3 \, \frac{(s-u+t)}{\lambda^{3/2}(s,u,t)}\,\, .
[m_c^2 (m_c^2 - s - u + t) + su] ~.
\eqa
The light quark condensate contributions ($B_{\mu}$) vanishes after double Borel transformation.  

In the phenomenological side, the correlator
(after double Borel transformation) is given by
\begin{eqnarray}
&&\B_{P^2}\left[\B_{P^{\prime 2}} 
[ \Gamma_{\mu}^{(\omega)} |_{phen}] \right]=
- \left( \frac{m_D^2 f_D}{m_c} \right)^2 \,m_{\omega}\,f_{\omega}
\,\frac{g_{\omega DD}^{(\omega)}(Q^2)}{m_\omega^2 + Q^2}\nonumber\\
&&\times e^{-m_D^2/M^2}e^{-m_D^2/M^{\prime 2}}
\left( \pp_{\mu} + p_{\mu} \right)
+ \B_{P^2}\left[\B_{P^{\prime 2}} [\mbox{HR}] \right] \, .
\label{eq:pheno2}
\end{eqnarray}
Now it is straightforward to construct the sum rules and
to calculate the value of vertex form factor $g_{\omega DD}$. 

{\bf Decay constants and thresholds.}
In the numerical evaluation of sum rules one can take into account 
the dependence of the decay constants with the Borel mass. In general
this procedure improves the stability of the vertex sum rules with 
respect to the increasing of Borel mass. From the sum rules for the 
correlators of the mesons we obtained the decay constants as 
functions of the Borel mass:
\begin{eqnarray}
f_{\omega}^2 (M^2) &=& \left[ \frac{M^2}{4\pi^2} (1 - e^{-s_0/M^2})
+ 2\frac{m_q\qq}{M^2} \right]  e^{m_{\omega}^2/M^2} ~,
\label{eq:fw2} \\
f_{D}^2 (M^2) &=& \frac{m_c^2}{m_D^4}
\left[ \frac{3}{8\pi^2} \int_{m_c^2}^{s_0} ds \frac{(s-m_c^2)^2}{s}
e^{-s/M^2} - m_c\qq e^{-m_c^2/M^2} \right] e^{m_{D}^2/M^2}
\label{eq:fD2}
\end{eqnarray}

As a common practice one usually assumes for the continuum 
threshold: $s_0 = (m_r + \Delta_r)^2$, where $m_r$ is the mass
of the ground state resonance and $\Delta_r \sim 0.5 - 0.6$GeV.  
For consistency we verified that the value of $\Delta_r$ above 
is suitable, at least for $\omega$ and $D$ mesons. Such verification 
can be done once $f_r (M^2)$ (actually $f_r(M^2, s_0)$)
stabilizes fast enough with growth of
$M^2$. Then we assume that $f_r(M^2 \rw \infty, s_0) = 
f_r(s_0)$ tends to the phenomenological (experimental) 
value of $f_r$.

For instance, in the case of the $\omega$ meson we can observe
from (\ref{eq:fw2}):
\beq
f_{\omega}^2 (M^2 \rw \infty, s_0) \rw \frac{s_0}{(2\pi)^2}
= \frac{(m_{\omega} + \Delta_{\omega})^2}{(2\pi)^2} \,\, ,
\eeq 
assuming $f_{\omega}^{(exp)} \approx 0.20$GeV and 
$m_{\omega} \approx 0.78$GeV, we obtain 
$\Delta_{\omega} \approx 0.48$GeV.

Repeating a similar calculation for eq.~(\ref{eq:fD2}) with:
$f_D \approx 0.17$GeV,  $m_{D} \approx 1.9$GeV, 
$m_{c} \approx 1.2$GeV and $\qq \approx (-0.23$GeV$)^3$, we
estimate $ \Delta_D \approx 0.53$GeV. We have seen that 
$\Delta_r \approx 0.6$GeV led to better stability of 
sum rules for larger values of Borel mass.

\section{Numerical results and discussions}
\noindent

As one can observe, in the sum rules above we have two structures:
one related to $p_{\mu}$ and other related to $\pp_{\mu}$.
For $D$ off-shell the last structure is more stable 
(with respect to Borel mass) than the first one \footnote{
On the other hand, for $\omega$ off--shell one can see that both 
structures are equal.}.
Therefore we concentrated our efforts in the analysis of $\pp_{\mu}$
structure.

The first step of the numerical analysis is to consider the
behavior of the form factor vertex with respect to the variations of Borel 
masses.
One important point that appears in this step is how to handle with the 
Borel masses $M$ and $M^{\prime}$. It is more or less intuitive
that the Borel mass must be of the same order of magnitude of
the mass of the corresponding particle. Then, for $D$ off--shell,
we take the ratio $M/M^{\prime} \approx m_{\omega}/m_D$, and for 
$\omega$ off--shell we have $M/M^{\prime} \approx m_D/m_D $.

\begin{figure}
{\centering 
\resizebox*{0.33\textheight}{!}
{\includegraphics{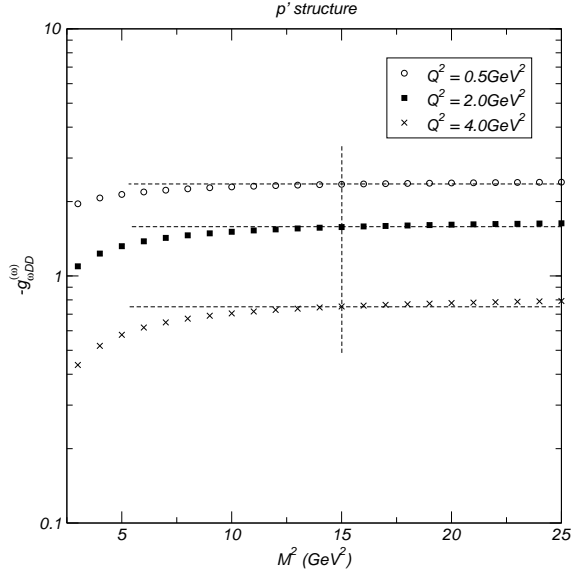}} 
\par}
\caption{
Behavior of $(- g_{\omega DD}^{(\omega)})$ with $M^2$ for
some fixed values of $Q^2$.}
\label{fig:vtx_M2}
\end{figure}

In fig.~\ref{fig:vtx_M2} we show the curves corresponding to
$g_{\omega DD}^{(\omega)}$ as function of Borel mass for
some fixed values of $Q^2$. We can note that $g_{\omega DD}^{(\omega)}$
stabilizes for $M^2 \gtrsim 15$GeV$^2$. A similar analysis is performed for 
$D$ off--shell, and we noted that $g_{\omega DD}^{(D)}$ stabilizes for
$M^2 \gtrsim 10$GeV$^2$.

\begin{figure}
{\centering 
\resizebox*{0.33\textheight}{!}
{\includegraphics{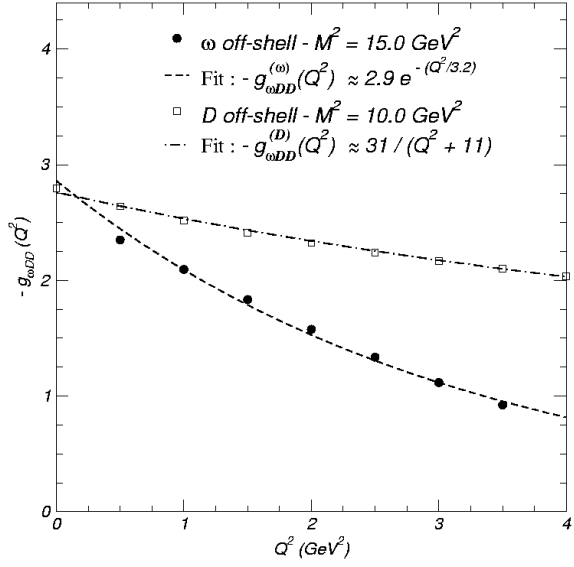}} 
\par}
\caption{
Behavior of $(- g_{\omega DD}(Q^2))$ for $\omega$ and $D$
off--shell ($p^\prime$ structure).}
\label{fig:vtx_Q2}
\end{figure}

The second step is to fix the Borel mass in values for which
the vertex form factor stabilizes, and study its dependence with
$Q^2$. In figure \ref{fig:vtx_Q2} we plot our results for both
situations: $D$ off--shell with $M^2 = 10$GeV$^2$ ($\Box$) 
and $\omega$ off--shell $M^2 = 15$GeV$^2$ ($\bullet$).

We fitted some functions to the curves of figure \ref{fig:vtx_Q2}.
For $\omega$ off--shell we found:
\beq
 - g_{\omega DD}^{(\omega)} (Q^2) \approx 2.9\, e^{- Q^2/3.2} \, ,
\eeq
and for $D$ off--shell
\beq
 - g_{\omega DD}^{(D)} (Q^2) \approx \frac{31}{Q^2 + 11} \, .
\eeq

From these fits we can estimate  $g_{\omega DD} \approx -2.9$
at zero momentum transfer, which perfectly agrees with the result of VMD 
model.  

In conclusion, we should say that the use of QCD Sum Rules approach to evaluate the
$g_{\omega DD}$ vertex is completely acceptable and demostrates that it is still an
open way to investigate other important hadronic quantities related with the formation
of the quark-gluon plasma. 

\section*{Acknowledgements}

We thank CAPES (LBH) and FAPESP (AM and RSMC) for financial support. 
RSMC thanks T. Frederico for useful suggestions and UNESP - Itapeva for 
hospitality. AM thanks M. Chiapparini for discussions. 



\begin{thebibliography}{99}

\bibitem{matsui} T. Matsui, H. Satz, Phys.\ Lett.\ {\bf B 187} (1986) 416; 
R. Vogt, Phys.\ Rep.\ {\bf 310} (1999) 197.

\bibitem{muller} For recent review see B. M\"uller {\tt nucl-th/0508062}.

\bibitem{finpe} See for instance: R.\ S.\ Azevedo, F.\ S.\ Navarra and M.\ Nielsen,
Acta Phys.\ Hung.~{\bf A 24} (2005) 253; R.\ S.\ Azevedo, M.\ Nielsen, 
Phys.\ Rev.\ {\bf C 69} (2004) 035201. 

\bibitem{bianco} For experimental dicussion see S. Bianco {\tt hep-ex/0512073}.

\bibitem{muller1} S.\ G.\ Matinyan and B. M\"uller, Phys.\ Rev.\ {\bf C 58}
 (1998) 2994. 

\bibitem{haglin} K.\ Haglin, Phys.\ Rev.\ {\bf C 61} (2000) 031902. 

\bibitem{lin-ko} Z.\ Lin,  C.\ M.\ Ko, Phys.\ Rev.\ {\bf C 62} 
(2000) 034903.

\bibitem{klingl} F.\ Klingl, N.\ Kaiser and W.\ Weise, Z.\ Phys.\
{\bf A 356} (1996) 193.

\bibitem{svz} M.\ A.\ Shifman, A.\ I.\ Vainshtein, V.\ I.\ Zakharov, 
Nucl.\ Phys.\ {\bf B 147} (1979) 385; Nucl. Phys. B 147 (1979) 448.

\bibitem{yang}
K.-C.\ Yang, W.-Y.\ P.\ Hwang, E.\ M.\  Henley, L.\ S.\  Kisslinger, 
Phys.\ Rev.\ {\bf D 47} (1993) 3001.

\bibitem{reinders}
H.\ Rubinstein, S.\ Yazaki, L.\ J.\  Reinders, 
Phys.\ Rep.\ {\bf 127} (1985) 1.

\bibitem{ioffe1}
B.\ L.\ Ioffe and A.\ V.\ Smilga, Nucl.\ Phys.\ {\bf B 216} (1983) 373.

\bibitem{ioffe2}
B.\ L.\ Ioffe and A.\ V.\ Smilga\ Nucl.\ Phys.\ {\bf B 232} (1984) 109.

\bibitem{revqcdsr} P.\ Colangelo, A.\ Khodjamirian, {\tt hep-ph/0010175}.

\bibitem{jpsi} R.~D.~Matheus, F.~S.~Navarra, M.~Nielsen and 
R.~Rodrigues da Silva, 
Phys.\ Lett.\ {\bf B 541}(2002) 265.

\bibitem{rho}M.~E.~Bracco, M.~Chiapparini, A.~Lozea, F.~S.~Navarra and 
M.~Nielsen, 
Phys.\ Lett.\ {\bf B 521} (2001) 1.

\bibitem{muta} 
T.\ Muta,
\newblock {\em Foundations of Quantum Chromodynamics}, World Scientific,
Singapore, (1987).

\end{thebibliography}
\end{document}